\documentclass[sigconf]{acmart}

\copyrightyear{2025}
\acmYear{2025}
\setcopyright{rightsretained}
\acmConference[UMAP Adjunct '25]{Adjunct Proceedings of the 33rd ACM Conference on User Modeling, Adaptation and Personalization}{June 16--19, 2025}{New York City, NY, USA}
\acmBooktitle{Adjunct Proceedings of the 33rd ACM Conference on User Modeling, Adaptation and Personalization (UMAP Adjunct '25), June 16--19, 2025, New York City, NY, USA}
\acmDOI{10.1145/3708319.3733687}
\acmISBN{979-8-4007-1399-6/2025/06}

\def\plainauthor{Kayhan Latifzadeh, Luis A. Leiva}
\def\plaintitle{Thalamus: A User Simulation Toolkit for Prototyping Multimodal Sensing Studies}
\def\plainkeywords{User Simulation; Behavioral/Physiological Sensing}

\usepackage[utf8]{inputenc}
\usepackage{wasysym}
\usepackage{bm}
\usepackage{url}
\usepackage{subfig}
\usepackage{colortbl}
\usepackage{booktabs}
\usepackage{multirow}
\usepackage{tablefootnote}
\usepackage[bottom]{footmisc}
\usepackage{enumitem}
\usepackage{soul} 

\graphicspath{{./img/}{./figs/},{./plots/}}

\hypersetup{%
  pdftitle={\plaintitle},
  pdfauthor={\plainauthor},
  pdfkeywords={\plainkeywords},
  bookmarksnumbered,
  pdfstartview={FitH},
  colorlinks,
  allcolors=black,
  breaklinks,
}

\begin{document}

\title[Thalamus Simulation Toolkit]{
    Thalamus: A User Simulation Toolkit\texorpdfstring{\\}{ }
    for Prototyping Multimodal Sensing Studies
}

\author{Kayhan Latifzadeh}
	\orcid{0000-0001-6172-0560}
	\affiliation{%
		\institution{University of Luxembourg}
		\city{Luxembourg}
		\country{Luxembourg}
        }
	\email{kayhan.latifzadeh@uni.lu}

	\author{Luis A. Leiva}
	\orcid{0000-0002-5011-1847}
	\affiliation{%
		\institution{University of Luxembourg}
		\city{Luxembourg}
		\country{Luxembourg}
        }
	\email{luis.leiva@uni.lu}

\renewcommand{\shortauthors}{K. Latifzadeh and L.A. Leiva}

\begin{abstract}
Conducting user studies that involve physiological and behavioral measurements is very time-consuming and expensive,
as it not only involves a careful experiment design, device calibration, etc.
but also a careful software testing.
We propose Thalamus, a software toolkit for collecting and simulating multimodal signals 
that can help the experimenters to prepare in advance for unexpected situations 
before reaching out to the actual study participants
and even before having to install or purchase a specific device.
Among other features, Thalamus allows the experimenter to modify, synchronize, and broadcast physiological signals
(as coming from various data streams) from different devices simultaneously
and not necessarily located in the same place.
Thalamus is cross-platform, cross-device, and simple to use, 
making it thus a valuable asset for HCI research.
\end{abstract}



\maketitle


\section{Introduction and Related Work}

In HCI research and experimentation, gathering multimodal information about the user 
can be crucial for comprehending the complex and dynamic nature of user interaction~\cite{lopes2021physiological}.
By simultaneously gathering different sensory data,
researchers can capture a more complete and nuanced picture of user behavior,
and investigate how various modalities interact and influence each other. 
For instance, monitoring eye movements together with facial expressions and brain activity
will provide valuable information about how users interpret and react to visual stimuli,
and can reveal patterns and trends that might not be apparent when focusing on just one modality alone~\cite{shoumy2020multimodal}.
As it enables researchers to comprehend how people use a system~\cite{peng2022detecting, kim2017sensors, dillen2020keep}, 
simultaneously gathering multimodal information may help design and evaluate novel technologies and interfaces. 

On the other hand, collecting multimodal signals is challenging, 
particularly when recording happens in real-time~\cite{sun2020insma}. 
It is essential for researchers to ensure that everything works correctly before conducting the main experiment,
otherwise they may incur in unforeseen costs derived e.g. from lab setup, participant payment, data post-processing, etc. 
A beneficial solution to address these issues is simulating the experiment before actually running it.
This way, researchers can ensure that the equipment and methods used for data collection 
are capable of handling the data streams they want to consider, 
that the data is of good quality, and that the system being tested can handle corner cases
such as device disconnections, missing or noisy data being recorded, etc.

Furthermore, data simulation can be used to test and optimize multimodal signal processing methods, 
evaluate the effectiveness of data analysis, identify potential sources of error, and, in short, 
save valuable time and resources by identifying and resolving potential problems early on.
In a nutshell, simulation is a useful and viable method for research in HCI, 
as it can be utilized in multiple ways to examine human behavior with technology~\cite{murray2022simulation},
before actual construction through prototyping~\cite{riegler2019autowsd},
offering training in a secure and controlled setting for individuals to utilize new technology.

To the best of our knowledge, there is currently no widely used software 
that is specifically designed for simulating different data streams in HCI experiments.
When it comes to recording multiple signal types simultaneously, 
many researchers have had to rely on ad-hoc solutions~\cite{kimchi2020opbox, notaro2018simultaneous, zimmerman2009observer, shah2020anxiety, xue2017crucial, ragot2017emotion, xiao2022time, wolling2021ibsync, szajerman2018joint, boekgaard2014twinkling, taib2014synchronising}, 
which can be time-consuming and difficult to replicate~\cite{latifzadeh2022gustav}. 
Moreover, previous work has proposed solutions for collecting multimodal signals, not for simulating them.
While ad-hoc solutions can solve a specific problem easily, 
they are not as generalizable as a dedicated software toolkit, 
and they may not provide the same level of consistency and repeatability across different experiment setups.

To address this gap, we propose Thalamus,
a toolkit for capturing and simulating various data streams, 
without the need to recruit users at the early stages of design, 
and even without the need to install or purchase a specific device. 
Among other features, Thalamus allows the experimenter to modify, synchronize, and broadcast physiological signals
(as coming from various data streams) from different devices simultaneously
and not necessarily located in the same place.
Thalamus is cross-platform, cross-device, and simple to use, 
making it thus a valuable asset for HCI research.

\section{The Thalamus toolkit}

\begin{figure}[!ht]
  \centering
  \includegraphics[width=\linewidth]{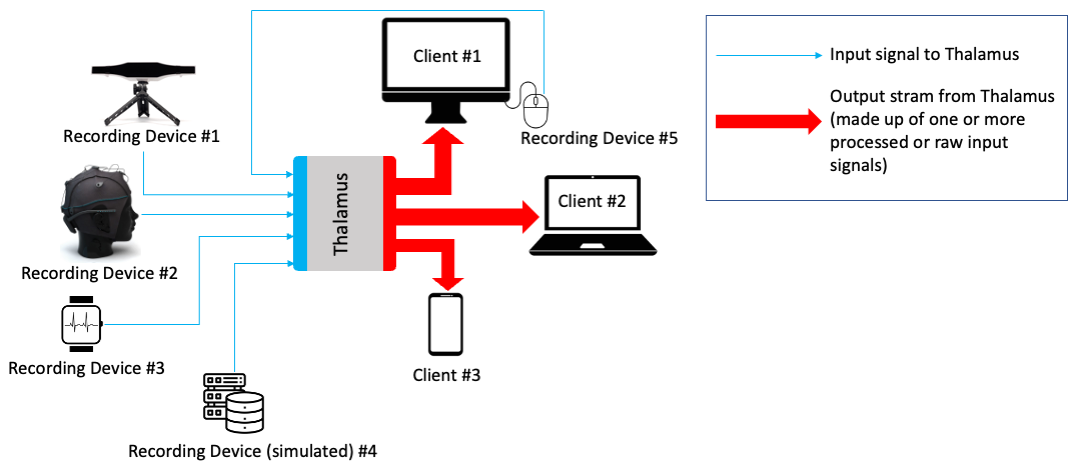}
  \caption{
    Conceptual diagram of the Thalamus toolkit,
    illustrating how it can be used to simulate different types of recording devices 
    and synchronize various data streams, including e.g. brain signals and eye movements. 
    Thalamus can receive feedback from any device. 
    For example, here client \#1 can also act as a recording device (Recording Device \#5), 
    thereby enabling real-time multimodal data collection and analysis.
  }

  \label{fig:architecture}
\end{figure}

Our proposed toolkit, depicted in \autoref{fig:architecture},
consists of three key components: 
a central hub for signal input/output, called \textit{Thalamus Core}, 
a number of \textit{Recording devices} that can be either simulated or real, 
and a number of \textit{Clients} that can receive one or more processed or raw sensory signals. 
Taken together, these components demonstrate how the toolkit can be utilized 
to simulate and coordinate a variety of data streams, 
including diverse types of recording devices.
Thalamus is written in Python and will be publicly available upon the publication of this paper.

\subsection{Thalamus Core}
The central component of our proposed toolkit is named after the Thalamus, 
a part of the brain that acts as a hub for processing and relaying sensory information to other areas. 
Conveniently, Thalamus Core acts as the hub for coordinating and synchronizing data streams, 
both from real and simulated signal recording devices. 
It is responsible for receiving data streams from various sources, processing and organizing it, 
and then streaming them to the clients for computation and analysis. 
This component is implemented using socket programming over the TCP/IP 
communication protocol,
which enables seamless communication and data transfer between the Thalamus, clients, and devices, 
ensuring efficient and real-time data stream coordination and synchronization.

\subsection{Recording devices}
In our proposed toolkit, recording devices can be real (i.e. physical devices that capture and record data) 
or simulated (i.e. virtual devices that mimic the functionality of the physical recording devices). 
Simulated devices are `injected' from public datasets through a controller
that parses the data and formats it to the expected transport format, based on JSON. 
We suggest JSON for data communication in Thalamus since it is a popular format that offers key advantages, 
including its self-describing nature, data types, cross-platform compatibility, and compact size. 
Any device, whether virtual or physical, 
is required to supply a coordinated universal time (UTC) timestamp for each recorded sample.
This is crucial for coordinating various signals and extracting events.
The use of UTC timestamps guarantees that data from various devices is collected and processed in a uniform and standardized manner, making it simple to compare and analyze.

Integrating real-world devices into our proposed software toolkit entails two main challenges. 
First, some devices may not provide UTC timestamps, 
which is crucial for extracting events and synchronizing different signals. 
Second, some devices may not provide easy access to the data stream, 
either because they are proprietary or they rely on a non-documented exchange format.
To solve these challenges, the experimenter can write controllers for each device that can format the data accordingly. 
For example, many devices write data to a file, and by creating a custom controller, 
it is possible to access and read the data sample by sample and reformat it. 
Then, the reformatted data can then be sent to the Thalamus Core component for further processing.

\subsection{Clients}
Any client that can open a socket connection can communicate with Thalamus, 
regardless of the operating system or programming languages used. 
This feature makes it easy for many devices to be considered in an experiment, 
ranging e.g. from computers, smartphones, and IoT devices. 
It also allows for greater flexibility and adaptability to the researcher's needs, 
as they can select the devices that are most convenient for their specific study.

When a device is connected to Thalamus, 
the Core component will provide the device with a list of available signals that can be recorded and streamed. 
This allows the device to select and prioritize the signals that are relevant to the specific study. 
Then the client specifies the type of data they wish to receive 
and the communication process with the Thalamus is initiated. 

In addition to receiving data streams, our toolkit also allows clients to act as recording devices themselves
and send their signals back to the Thalamus Core. 
For example, in an experiment where researchers would like to share the mouse movements of one client with another, 
the mouse signal can be streamed back to the Thalamus Core and be broadcasted to the other clients. 

\subsection{Built-in features and functions}
Our simulation toolkit offers a variety of built-in functionalities, 
such as dealing with missing data, applying common filters, synchronizing multiple signals, simulating delays, and adding signal noise. We explain these key features in the following sections.

\subsubsection{Missing values:}
\begin{figure}[!ht]
  \centering
  \includegraphics[width=\linewidth]{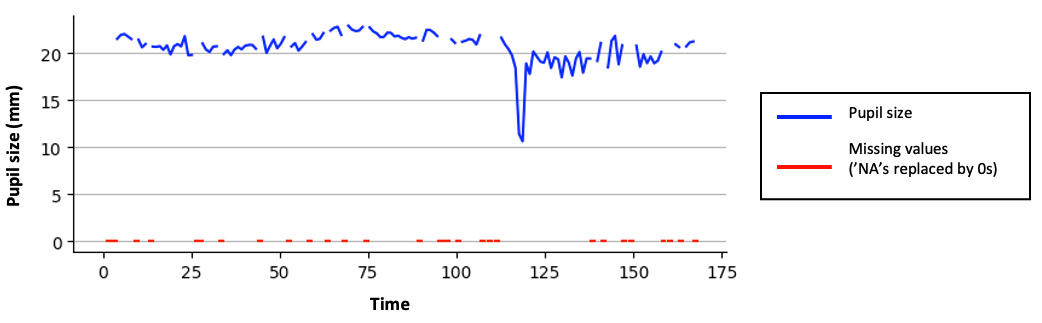}
  \caption{
    Demonstration of the toolkit's ability to handle missing values.
    This figure illustrates the pupil size values obtained from a sample of eye-tracking data. Missing values represented by \texttt{"NA"} are replaced with zeros.
  }

  \label{fig:features-missing}
\end{figure}

In a real-world setting, missing values are a common issue that can occur during data collection. 
For example, in eye-tracking recordings, if the participant looks outside the screen, 
there would be no valid value of eye movement to record. 
To simulate this issue, Thalamus sends a conventional value such as \texttt{"0"} or \texttt{"NA"} 
as an indicator of missing values (see \autoref{fig:features-missing}). 
This allows the clients to recognize and properly handle these missing values. 

\subsubsection{Filters:}
\begin{figure}[!ht]
  \centering
  \includegraphics[width=\linewidth]{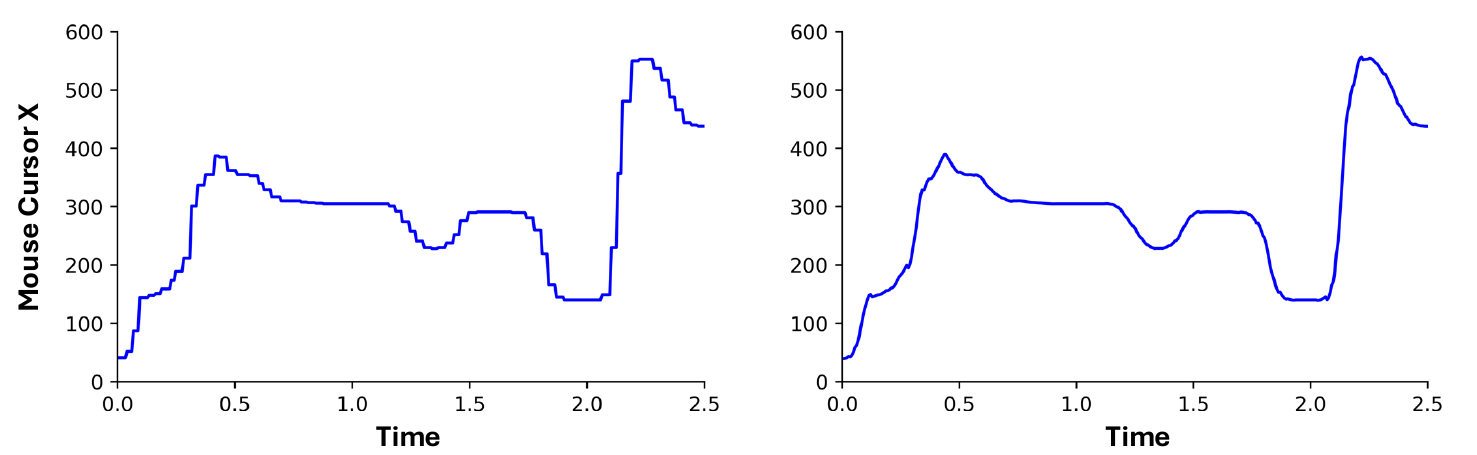}
  \caption{
    Demonstration of the toolkit's ability to provide common filters.
    An original signal, in this case the mouse cursor position in the X axis, is show on the left part. 
    The same signal after being filtered with the Savitzky-Golay filter is shown on the right.
  }

  \label{fig:features-filter}
\end{figure}

The ability to apply various kinds of filters to the data streams is another feature of our toolkit.
For example, currently Savitzky-Golay~\cite{savitzky1964smoothing} (see \autoref{fig:features-filter})
and Kalman~\cite{li2015kalman} filters can be used.
These filters can be used to enhance the quality of the signal
by reducing noise, boosting the signal-to-noise ratio.
Because of the ability to apply various filters based on the objectives of the study, 
this feature gives researchers greater flexibility and adaptability.

\subsubsection{Signal synchronization:}
The ability to synchronize various data streams based on timestamps is another crucial component of our toolkit.
This capability is essential for correctly and effectively syncing various data streams 
and extracting information that is pertinent from the data.
It enables researchers to compare and examine data from many sources, 
as well as to extract events from the data streams.
By ensuring that data from many devices is processed in a uniform and standardized manner (using UTC timestamps), 
it is simple to compare and evaluate data from various devices and during different time periods.

\begin{figure}[!ht]
  \centering
  \includegraphics[width=\linewidth]{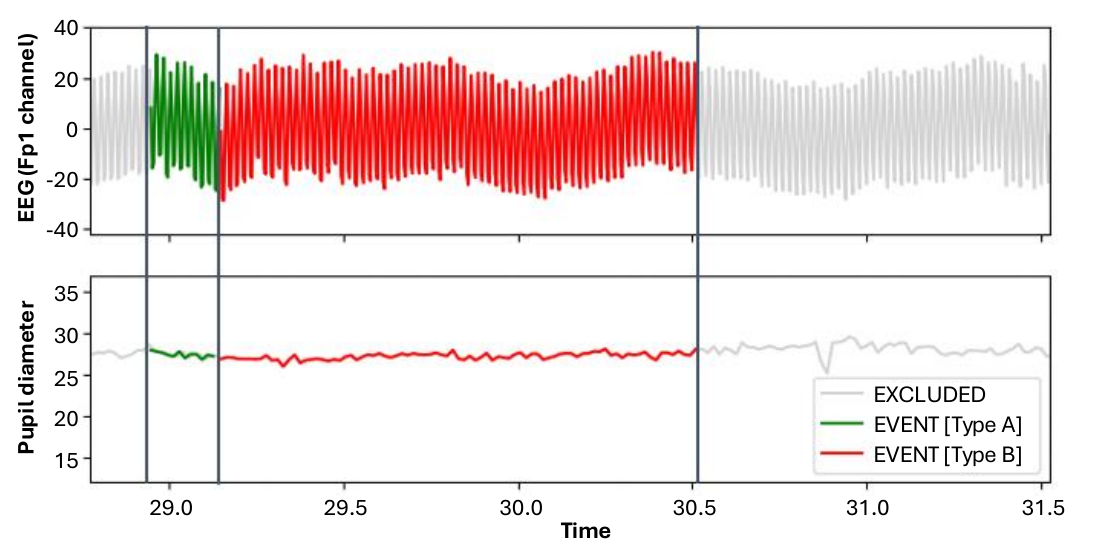}
  \caption{
    Demonstration of the toolkit's ability to synchronize simultaneous data streams.
    The upper timeseries represents the brain signal (EEG) 
    whereas the lower timeseries represents the pupil diameter. 
  }

  \label{fig:features-sync}
\end{figure}

From a technical perspective, all connected or simulated devices should have their system clocks in time
and submit their data associated with a Unix timestamp,
which is a standardized time reference --- the number of (milli)seconds elapsed since January 1, 1970. 
This way, Thalamus can synchronize all signals and extract specific time periods as needed.
For example, in an experiment that is collecting brain signals and pupil size, 
when presenting a stimulus to the participants, 
Thalamus can extract this specific time period and synchronize the pupil size signal with the brain signal.
\autoref{fig:features-sync} illustrates this capability.
The temporal precision in our proposed toolkit is in milliseconds, 
which is convenient for most studies in human-computer interaction.

\subsubsection{Noise:}
\begin{figure}[!ht]
  \centering
  \includegraphics[width=\linewidth]{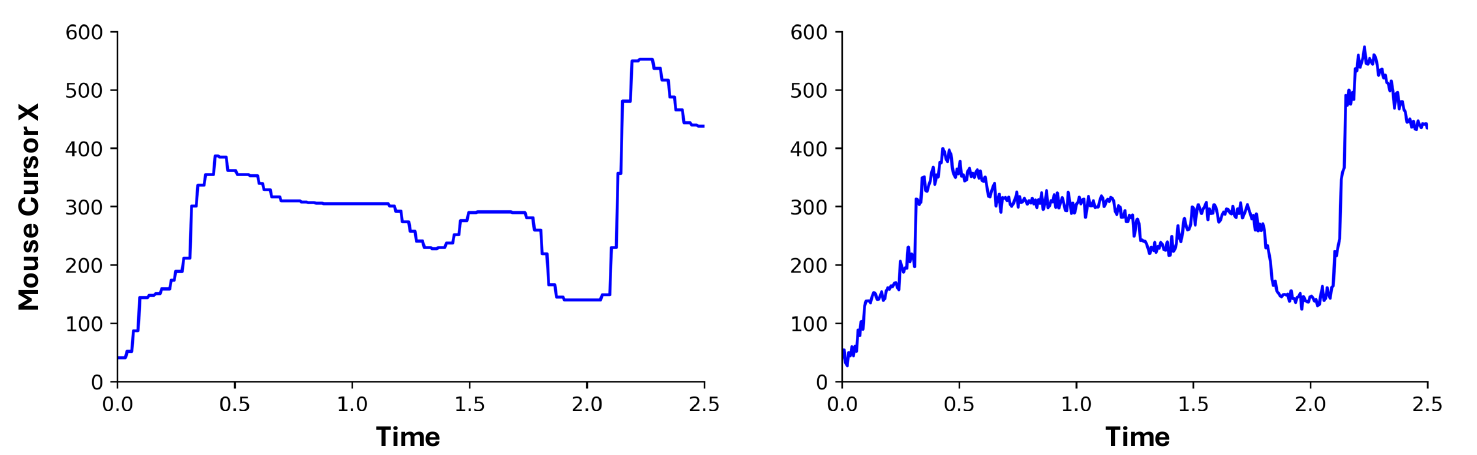}
  \caption{
    Demonstration of the toolkit's ability to simulate noise. 
    An original timeseries, in this case the mouse cursor position in the X axis, is shown on the left part. 
    The same signal is incorporated Gaussian noise, as shown on the right part.
  }

  \label{fig:features-noise}
\end{figure}

The presence of noise in the data streams is another pervasive issue that arises often in real-world circumstances.
To address this, our toolkit has the capacity to mimic several kinds of noise, 
including fixed (constant), random, and Gaussian noise (see \autoref{fig:features-noise}).
With the help of this feature, researchers may test and challenge their experimental setup in a controlled setting 
while accounting for any noise that may emerge during a genuine experiment using actual equipment. 

\subsubsection{Delay simulation:}
\begin{figure}[!ht]
  \centering
  \includegraphics[width=\linewidth]{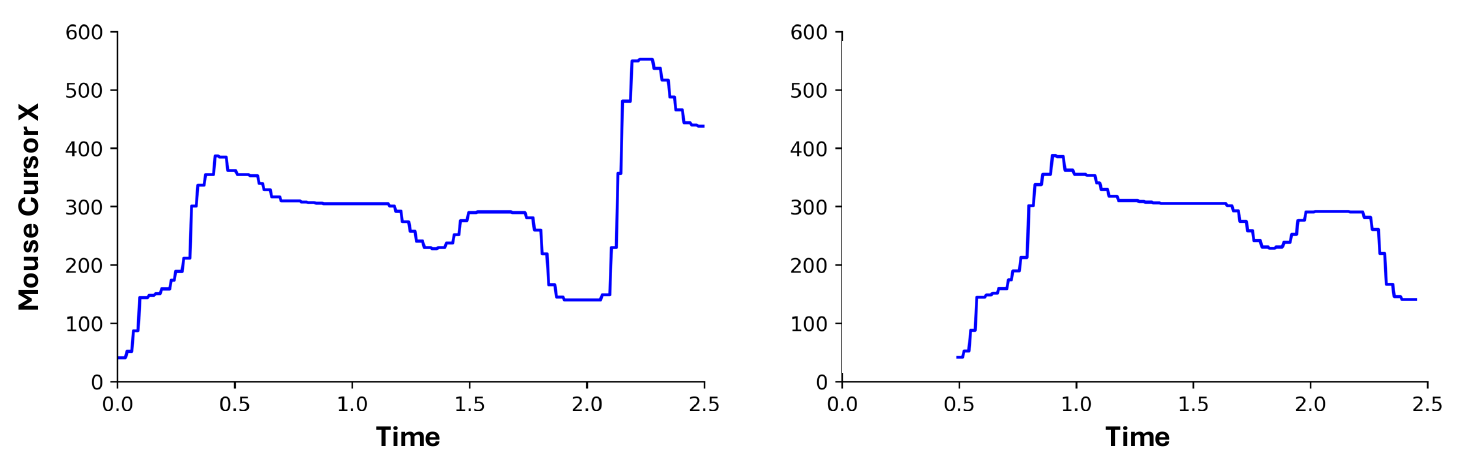}
  \caption{
    Demonstration of the toolkit's ability to simulate delayed signals. 
    An original timeseries, in this example the mouse cursor position in the X axis, is shown on the left part. 
    The same signal is incorporated a delay, as shown on the right part.
  }

  \label{fig:features-delay}
\end{figure}

Delays in receiving data streams may be problematic in real-world situations.
Our toolkit has the capacity to simulate delays, 
based on the timestamps or a buffer window (see \autoref{fig:features-delay}).
By simulating various delay scenarios and variations, 
this capability allows researchers to find and refine strategies for synchronizing and processing data streams, 
which can help to increase the system's robustness
by anticipating possible unstable connections.

\section{Use case examples}

In the following we illustrate a few use cases 
that allow for a comprehensive evaluation of Thalamus
in practical applications.

\subsection{Real-time applications}
\label{sec:rt}

A common scenario while streaming data in real-time application
is that delays may happen due to network issues or any other reason. 
Researchers can simulate these delays using the built-in features of Thalamus 
to troubleshoot such suspected situations. 
Also, common filtering and preprocessing procedures can be applied. 
This step is conducted before streaming the data to clients, 
resulting in a significant reduction in computational costs. 
By implementing these procedures once and transmitting the processed data to all clients, 
Thalamus enhances efficiency and optimizes resource utilization for researchers.

\subsection{Try before you buy}
\label{sec:try}

A research aims to conduct a study 
using Brain-Computer Interfaces (BCIs) to measure the emotional responses of people while watching video clips. 
Since there is a variety of BCIs in the market to collect EEG signals, 
each with a different number of channels,
the researcher plans to utilize publicly available datasets such as 
DREAMER~\cite{katsigiannis2017dreamer}, MAHNOB~\cite{soleymani2011multimodal}, and SEED~\cite{zheng2015investigating},
which are specifically focused on emotional responses to videos using EEG signals, for the simulation. 
The researcher uses Thalamus as a cost-effective test method
to decide if they should purchase a 62-channel EEG device (like in SEED)
or if a 14-channel device (like in DREAMER) would suffice.

\subsection{Device stress-testing}
\label{sec:stress}

A group of researchers are planning to conduct an experiment to measure cognitive load, utilizing various modalities such as eye-tracking, ECG, mouse movements, and skin temperature. They have all the necessary recording devices but they do not know their operational ranges,
so they use Thalamus to test a few recordings with each device. This approach also allows them to optimize their data processing methods, ensure their pipeline works as expected, and select the most suitable device for their experiment, all while saving time and resources.

\subsection{Remote data collection}
\label{sec:remote}

A researcher wants to develop an application
that sends an alarm when high cognitive load is detected 
during online lectures, where attendees are wearing physiological sensors. 
In this scenario, content is streamed to all attendees through the Internet, 
who may be located in different places. 
Before creating the application,
the researcher uses Thalamus to simulate multiple instances of different sensors
using public datasets, as demonstrated in \autoref{sec:try},
and troubleshoot challenges like external noise,
that could potentially affect the accuracy of some of the signals.

\subsection{Broadcasting signals}
\label{sec:sharing}

A medical researcher is planning to conduct an experiment on telemedicine, in which a group of cardiologists will provide their opinions on the status of a patient. They should all receive the exact same ECG stream on their devices (e.g., computers, tablets, etc.) in real-time. The researcher uses Thalamus to stream a batch of pre-recorded, partitioned, and status-labeled ECG data to multiple devices, allowing the researcher to assess the efficiency of the designed experiment.

\section{Discussion, Limitations, and Future Work}

We have introduced Thalamus, a toolkit for simulating, capturing, and manipulating different data streams. 
Thalamus allows researchers to conduct a dry-run of their experiments, 
to test their equipment and setup before running the actual experiment with real users,
which allows them to save valuable resources and expenses. 
Our toolkit includes features such as synchronizing multimodal signals, noise injection, delay simulations, and more. 
In sum, Thalamus provides researchers with a practical and streamlined method for simulating different data modalities.

One limitation of Thalamus is that it may not be able to simulate 
all types of data streams or devices one may think of ``out of the box''. 
It currently supports device simulation from structured files such as CSV or JSON.
Future work will also consider relational databases such as MySQL or Postgres.
Another limitation worth of mentioning is that 
programming skills may be needed to write custom controllers for some devices, 
for example those that do not provide direct access to the captured sensor data.

We also plan for future work to add more built-in functions to the Thalamus Core,
to alleviate the need to write them if they are commonly used by researchers.
Since our software is open source, we welcome contributions from the community. 
We also plan to develop a graphical user interface program 
that will allow the experimenter to configure our toolkit 
in a visual environment, similar to the popular Max/MSP software~\cite{Momeni03}.
Thalamus is publicly available at \url{https://github.com/kayhan-latifzadeh/Thalamus}.

\begin{acks}
This work is supported by the Horizon 2020 FET program of the European Union 
through the ERA-NET Cofund funding (BANANA, grant CHIST-ERA-20-BCI-001)
and Horizon Europe's European Innovation Council 
through the Pathfinder program (SYMBIOTIK, grant 101071147).
\end{acks}



\begin{thebibliography}{26}


\ifx \showCODEN    \undefined \def \showCODEN     #1{\unskip}     \fi
\ifx \showDOI      \undefined \def \showDOI       #1{#1}\fi
\ifx \showISBNx    \undefined \def \showISBNx     #1{\unskip}     \fi
\ifx \showISBNxiii \undefined \def \showISBNxiii  #1{\unskip}     \fi
\ifx \showISSN     \undefined \def \showISSN      #1{\unskip}     \fi
\ifx \showLCCN     \undefined \def \showLCCN      #1{\unskip}     \fi
\ifx \shownote     \undefined \def \shownote      #1{#1}          \fi
\ifx \showarticletitle \undefined \def \showarticletitle #1{#1}   \fi
\ifx \showURL      \undefined \def \showURL       {\relax}        \fi
\providecommand\bibfield[2]{#2}
\providecommand\bibinfo[2]{#2}
\providecommand\natexlab[1]{#1}
\providecommand\showeprint[2][]{arXiv:#2}

\bibitem[\protect\citeauthoryear{B{\oe}kgaard, Petersen, and Larsen}{B{\oe}kgaard et~al\mbox{.}}{2014}]%
        {boekgaard2014twinkling}
\bibfield{author}{\bibinfo{person}{P. B{\oe}kgaard}, \bibinfo{person}{M.~K. Petersen}, {and} \bibinfo{person}{J.~E. Larsen}.} \bibinfo{year}{2014}\natexlab{}.
\newblock \showarticletitle{In the twinkling of an eye: Synchronization of {EEG} and eye tracking based on blink signatures}. In \bibinfo{booktitle}{\emph{Proc. Intl. Workshop on Cognitive Information Processing (CIP)}}.
\newblock


\bibitem[\protect\citeauthoryear{Dillen, Ilievski, Law, Nacke, Czarnecki, and Schneider}{Dillen et~al\mbox{.}}{2020}]%
        {dillen2020keep}
\bibfield{author}{\bibinfo{person}{N. Dillen}, \bibinfo{person}{M. Ilievski}, \bibinfo{person}{E. Law}, \bibinfo{person}{L.~E. Nacke}, \bibinfo{person}{K. Czarnecki}, {and} \bibinfo{person}{O. Schneider}.} \bibinfo{year}{2020}\natexlab{}.
\newblock \showarticletitle{Keep calm and ride along: Passenger comfort and anxiety as physiological responses to autonomous driving styles}. In \bibinfo{booktitle}{\emph{Proceedings of the 2020 CHI conference on human factors in computing systems}}. \bibinfo{pages}{1--13}.
\newblock


\bibitem[\protect\citeauthoryear{Katsigiannis and Ramzan}{Katsigiannis and Ramzan}{2017}]%
        {katsigiannis2017dreamer}
\bibfield{author}{\bibinfo{person}{S. Katsigiannis} {and} \bibinfo{person}{N. Ramzan}.} \bibinfo{year}{2017}\natexlab{}.
\newblock \showarticletitle{{DREAMER}: A database for emotion recognition through {EEG} and {ECG} signals from wireless low-cost off-the-shelf devices}.
\newblock \bibinfo{journal}{\emph{IEEE journal of biomedical and health informatics}} \bibinfo{volume}{22}, \bibinfo{number}{1} (\bibinfo{year}{2017}), \bibinfo{pages}{98--107}.
\newblock


\bibitem[\protect\citeauthoryear{Kim, Patra, Kim, Lee, Segev, and Lee}{Kim et~al\mbox{.}}{2017}]%
        {kim2017sensors}
\bibfield{author}{\bibinfo{person}{S. Kim}, \bibinfo{person}{K.~A.~E. Patra}, \bibinfo{person}{A. Kim}, \bibinfo{person}{K.-P. Lee}, \bibinfo{person}{A. Segev}, {and} \bibinfo{person}{U. Lee}.} \bibinfo{year}{2017}\natexlab{}.
\newblock \showarticletitle{Sensors know which photos are memorable}. In \bibinfo{booktitle}{\emph{Proceedings of the 2017 CHI Conference Extended Abstracts on Human Factors in Computing Systems}}. \bibinfo{pages}{2706--2713}.
\newblock


\bibitem[\protect\citeauthoryear{Kimchi, Coughlin, Shanahan, Piantoni, Pezaris, and Cash}{Kimchi et~al\mbox{.}}{2020}]%
        {kimchi2020opbox}
\bibfield{author}{\bibinfo{person}{E.~Y. Kimchi}, \bibinfo{person}{B.~F. Coughlin}, \bibinfo{person}{B.~E. Shanahan}, \bibinfo{person}{G. Piantoni}, \bibinfo{person}{J. Pezaris}, {and} \bibinfo{person}{S.~S. Cash}.} \bibinfo{year}{2020}\natexlab{}.
\newblock \showarticletitle{{OpBox}: Open source tools for simultaneous {EEG} and {EMG} acquisition from multiple subjects}.
\newblock \bibinfo{journal}{\emph{eNeuro}} \bibinfo{volume}{7}, \bibinfo{number}{5} (\bibinfo{year}{2020}).
\newblock


\bibitem[\protect\citeauthoryear{Latifzadeh and Leiva}{Latifzadeh and Leiva}{2022}]%
        {latifzadeh2022gustav}
\bibfield{author}{\bibinfo{person}{K. Latifzadeh} {and} \bibinfo{person}{L.~A. Leiva}.} \bibinfo{year}{2022}\natexlab{}.
\newblock \showarticletitle{Gustav: Cross-device Cross-computer Synchronization of Sensory Signals}. In \bibinfo{booktitle}{\emph{The Adjunct Publication of the 35th Annual ACM Symposium on User Interface Software and Technology}}. \bibinfo{pages}{1--3}.
\newblock


\bibitem[\protect\citeauthoryear{Li, Li, Ji, and Dai}{Li et~al\mbox{.}}{2015}]%
        {li2015kalman}
\bibfield{author}{\bibinfo{person}{Q. Li}, \bibinfo{person}{R. Li}, \bibinfo{person}{K. Ji}, {and} \bibinfo{person}{W. Dai}.} \bibinfo{year}{2015}\natexlab{}.
\newblock \showarticletitle{Kalman filter and its application}. In \bibinfo{booktitle}{\emph{2015 8th International Conference on Intelligent Networks and Intelligent Systems (ICINIS)}}. IEEE, \bibinfo{pages}{74--77}.
\newblock


\bibitem[\protect\citeauthoryear{Lopes, Chuang, and Maes}{Lopes et~al\mbox{.}}{2021}]%
        {lopes2021physiological}
\bibfield{author}{\bibinfo{person}{P. Lopes}, \bibinfo{person}{L.~L. Chuang}, {and} \bibinfo{person}{P. Maes}.} \bibinfo{year}{2021}\natexlab{}.
\newblock \showarticletitle{Physiological {I/O}}. In \bibinfo{booktitle}{\emph{Extended Abstracts of the 2021 CHI Conference on Human Factors in Computing Systems}}. \bibinfo{pages}{1--4}.
\newblock


\bibitem[\protect\citeauthoryear{Momeni and Wessel}{Momeni and Wessel}{2003}]%
        {Momeni03}
\bibfield{author}{\bibinfo{person}{A. Momeni} {and} \bibinfo{person}{D. Wessel}.} \bibinfo{year}{2003}\natexlab{}.
\newblock \showarticletitle{Characterizing and Controlling Musical Material Intuitively with Graphical Models}. In \bibinfo{booktitle}{\emph{Proceedings of the New Interfaces for Musical Expression Conference}}.
\newblock


\bibitem[\protect\citeauthoryear{Murray-Smith, Oulasvirta, Howes, M{\"u}ller, Ikkala, Bachinski, Fleig, Fischer, and Klar}{Murray-Smith et~al\mbox{.}}{2022}]%
        {murray2022simulation}
\bibfield{author}{\bibinfo{person}{R. Murray-Smith}, \bibinfo{person}{A. Oulasvirta}, \bibinfo{person}{A. Howes}, \bibinfo{person}{J. M{\"u}ller}, \bibinfo{person}{A. Ikkala}, \bibinfo{person}{M. Bachinski}, \bibinfo{person}{A. Fleig}, \bibinfo{person}{F. Fischer}, {and} \bibinfo{person}{M. Klar}.} \bibinfo{year}{2022}\natexlab{}.
\newblock \showarticletitle{What simulation can do for HCI research}.
\newblock \bibinfo{journal}{\emph{Interactions}} \bibinfo{volume}{29}, \bibinfo{number}{6} (\bibinfo{year}{2022}), \bibinfo{pages}{48--53}.
\newblock


\bibitem[\protect\citeauthoryear{Notaro and Diamond}{Notaro and Diamond}{2018}]%
        {notaro2018simultaneous}
\bibfield{author}{\bibinfo{person}{G.~M. Notaro} {and} \bibinfo{person}{S.~G. Diamond}.} \bibinfo{year}{2018}\natexlab{}.
\newblock \showarticletitle{Simultaneous {EEG}, eye-tracking, behavioral, and screen-capture data during online German language learning}.
\newblock \bibinfo{journal}{\emph{Data Brief}}  \bibinfo{volume}{21} (\bibinfo{year}{2018}).
\newblock


\bibitem[\protect\citeauthoryear{Peng, Meng, Xie, Huang, Chen, and Wang}{Peng et~al\mbox{.}}{2022}]%
        {peng2022detecting}
\bibfield{author}{\bibinfo{person}{X. Peng}, \bibinfo{person}{C. Meng}, \bibinfo{person}{X. Xie}, \bibinfo{person}{J. Huang}, \bibinfo{person}{H. Chen}, {and} \bibinfo{person}{H. Wang}.} \bibinfo{year}{2022}\natexlab{}.
\newblock \showarticletitle{Detecting challenge from physiological signals: A primary study with a typical game scenario}. In \bibinfo{booktitle}{\emph{CHI Conference on Human Factors in Computing Systems Extended Abstracts}}. \bibinfo{pages}{1--7}.
\newblock


\bibitem[\protect\citeauthoryear{Ragot, Martin, Em, Pallamin, and Diverrez}{Ragot et~al\mbox{.}}{2017}]%
        {ragot2017emotion}
\bibfield{author}{\bibinfo{person}{M. Ragot}, \bibinfo{person}{N. Martin}, \bibinfo{person}{S. Em}, \bibinfo{person}{N. Pallamin}, {and} \bibinfo{person}{J.-M. Diverrez}.} \bibinfo{year}{2017}\natexlab{}.
\newblock \showarticletitle{Emotion Recognition Using Physiological Signals: Laboratory vs. Wearable Sensors}. In \bibinfo{booktitle}{\emph{Proc. Intl. Conf. Applied Human Factors and Ergonomics (AHFE)}}.
\newblock


\bibitem[\protect\citeauthoryear{Riegler, Riener, and Holzmann}{Riegler et~al\mbox{.}}{2019}]%
        {riegler2019autowsd}
\bibfield{author}{\bibinfo{person}{A. Riegler}, \bibinfo{person}{A. Riener}, {and} \bibinfo{person}{C. Holzmann}.} \bibinfo{year}{2019}\natexlab{}.
\newblock \showarticletitle{{AutoWSD}: Virtual reality automated driving simulator for rapid HCI prototyping}.
\newblock In \bibinfo{booktitle}{\emph{Proceedings of Mensch und Computer 2019}}. \bibinfo{pages}{853--857}.
\newblock


\bibitem[\protect\citeauthoryear{Savitzky and Golay}{Savitzky and Golay}{1964}]%
        {savitzky1964smoothing}
\bibfield{author}{\bibinfo{person}{A. Savitzky} {and} \bibinfo{person}{M.~J. Golay}.} \bibinfo{year}{1964}\natexlab{}.
\newblock \showarticletitle{Smoothing and differentiation of data by simplified least squares procedures.}
\newblock \bibinfo{journal}{\emph{Analytical chemistry}} \bibinfo{volume}{36}, \bibinfo{number}{8} (\bibinfo{year}{1964}), \bibinfo{pages}{1627--1639}.
\newblock


\bibitem[\protect\citeauthoryear{Shah, Chow, and Lee}{Shah et~al\mbox{.}}{2020}]%
        {shah2020anxiety}
\bibfield{author}{\bibinfo{person}{E.~J. Shah}, \bibinfo{person}{J.~Y. Chow}, {and} \bibinfo{person}{M.~J. Lee}.} \bibinfo{year}{2020}\natexlab{}.
\newblock \showarticletitle{Anxiety on Quiet Eye and Performance of Youth Pistol Shooters}.
\newblock \bibinfo{journal}{\emph{J. Sport Exerc. Psychol.}} \bibinfo{volume}{42}, \bibinfo{number}{4} (\bibinfo{year}{2020}).
\newblock


\bibitem[\protect\citeauthoryear{Shoumy, Ang, Seng, Rahaman, and Zia}{Shoumy et~al\mbox{.}}{2020}]%
        {shoumy2020multimodal}
\bibfield{author}{\bibinfo{person}{N.~J. Shoumy}, \bibinfo{person}{L.-M. Ang}, \bibinfo{person}{K.~P. Seng}, \bibinfo{person}{D.~M. Rahaman}, {and} \bibinfo{person}{T. Zia}.} \bibinfo{year}{2020}\natexlab{}.
\newblock \showarticletitle{Multimodal big data affective analytics: A comprehensive survey using text, audio, visual and physiological signals}.
\newblock \bibinfo{journal}{\emph{Journal of Network and Computer Applications}}  \bibinfo{volume}{149} (\bibinfo{year}{2020}), \bibinfo{pages}{102447}.
\newblock


\bibitem[\protect\citeauthoryear{Soleymani, Lichtenauer, Pun, and Pantic}{Soleymani et~al\mbox{.}}{2011}]%
        {soleymani2011multimodal}
\bibfield{author}{\bibinfo{person}{M. Soleymani}, \bibinfo{person}{J. Lichtenauer}, \bibinfo{person}{T. Pun}, {and} \bibinfo{person}{M. Pantic}.} \bibinfo{year}{2011}\natexlab{}.
\newblock \showarticletitle{A multimodal database for affect recognition and implicit tagging}.
\newblock \bibinfo{journal}{\emph{IEEE transactions on affective computing}} \bibinfo{volume}{3}, \bibinfo{number}{1} (\bibinfo{year}{2011}), \bibinfo{pages}{42--55}.
\newblock


\bibitem[\protect\citeauthoryear{Sun, Guo, Kaffashi, Jacono, DeGeorgia, and Loparo}{Sun et~al\mbox{.}}{2020}]%
        {sun2020insma}
\bibfield{author}{\bibinfo{person}{Y. Sun}, \bibinfo{person}{F. Guo}, \bibinfo{person}{F. Kaffashi}, \bibinfo{person}{F.~J. Jacono}, \bibinfo{person}{M. DeGeorgia}, {and} \bibinfo{person}{K.~A. Loparo}.} \bibinfo{year}{2020}\natexlab{}.
\newblock \showarticletitle{{INSMA}: An integrated system for multimodal data acquisition and analysis in the intensive care unit}.
\newblock \bibinfo{journal}{\emph{Journal of biomedical informatics}}  \bibinfo{volume}{106} (\bibinfo{year}{2020}), \bibinfo{pages}{103434}.
\newblock


\bibitem[\protect\citeauthoryear{Szajerman, Napieralski, and Lecointe}{Szajerman et~al\mbox{.}}{2018}]%
        {szajerman2018joint}
\bibfield{author}{\bibinfo{person}{D. Szajerman}, \bibinfo{person}{P. Napieralski}, {and} \bibinfo{person}{J.-P. Lecointe}.} \bibinfo{year}{2018}\natexlab{}.
\newblock \showarticletitle{Joint analysis of simultaneous {EEG} and eye tracking data for video images}. In \bibinfo{booktitle}{\emph{Proc. IEEE ISEF}}.
\newblock


\bibitem[\protect\citeauthoryear{Taib, Itzstein, and Yu}{Taib et~al\mbox{.}}{2014}]%
        {taib2014synchronising}
\bibfield{author}{\bibinfo{person}{R. Taib}, \bibinfo{person}{B. Itzstein}, {and} \bibinfo{person}{K. Yu}.} \bibinfo{year}{2014}\natexlab{}.
\newblock \showarticletitle{Synchronising Physiological and Behavioural Sensors in a Driving Simulator}. In \bibinfo{booktitle}{\emph{Proc. Intl. Conf. Multimodal Interaction (ICMI)}}.
\newblock


\bibitem[\protect\citeauthoryear{Wolling, Huynh, and Van~Laerhoven}{Wolling et~al\mbox{.}}{2021}]%
        {wolling2021ibsync}
\bibfield{author}{\bibinfo{person}{F. Wolling}, \bibinfo{person}{C.~D. Huynh}, {and} \bibinfo{person}{K. Van~Laerhoven}.} \bibinfo{year}{2021}\natexlab{}.
\newblock \showarticletitle{{IBSync}: Intra-body synchronization of wearable devices using artificial {ECG} landmarks}. In \bibinfo{booktitle}{\emph{Proc. Intl. Symposium on Wearable Computers (ISWC)}}.
\newblock


\bibitem[\protect\citeauthoryear{Xiao, Ding, and Hu}{Xiao et~al\mbox{.}}{2022}]%
        {xiao2022time}
\bibfield{author}{\bibinfo{person}{R. Xiao}, \bibinfo{person}{C. Ding}, {and} \bibinfo{person}{X. Hu}.} \bibinfo{year}{2022}\natexlab{}.
\newblock \showarticletitle{Time Synchronization of Multimodal Physiological Signals through Alignment of Common Signal Types and Its Technical Considerations in Digital Health}.
\newblock \bibinfo{journal}{\emph{J. Imaging}} \bibinfo{volume}{8}, \bibinfo{number}{5} (\bibinfo{year}{2022}).
\newblock


\bibitem[\protect\citeauthoryear{Xue, Quan, Li, Yue, and Zhang}{Xue et~al\mbox{.}}{2017}]%
        {xue2017crucial}
\bibfield{author}{\bibinfo{person}{J. Xue}, \bibinfo{person}{C. Quan}, \bibinfo{person}{C. Li}, \bibinfo{person}{J. Yue}, {and} \bibinfo{person}{C. Zhang}.} \bibinfo{year}{2017}\natexlab{}.
\newblock \showarticletitle{A crucial temporal accuracy test of combining {EEG} and {Tobii} eye tracker}.
\newblock \bibinfo{journal}{\emph{Medicine}} \bibinfo{volume}{96}, \bibinfo{number}{13} (\bibinfo{year}{2017}).
\newblock


\bibitem[\protect\citeauthoryear{Zheng and Lu}{Zheng and Lu}{2015}]%
        {zheng2015investigating}
\bibfield{author}{\bibinfo{person}{W.-L. Zheng} {and} \bibinfo{person}{B.-L. Lu}.} \bibinfo{year}{2015}\natexlab{}.
\newblock \showarticletitle{Investigating Critical Frequency Bands and Channels for {EEG}-based Emotion Recognition with Deep Neural Networks}.
\newblock \bibinfo{journal}{\emph{IEEE Transactions on Autonomous Mental Development}} \bibinfo{volume}{7}, \bibinfo{number}{3} (\bibinfo{year}{2015}).
\newblock
\urldef\tempurl%
\url{https://doi.org/10.1109/TAMD.2015.2431497}
\showDOI{\tempurl}


\bibitem[\protect\citeauthoryear{Zimmerman, Bolhuis, Willemsen, Meyer, and Noldus}{Zimmerman et~al\mbox{.}}{2009}]%
        {zimmerman2009observer}
\bibfield{author}{\bibinfo{person}{P.~H. Zimmerman}, \bibinfo{person}{J.~E. Bolhuis}, \bibinfo{person}{A. Willemsen}, \bibinfo{person}{E.~S. Meyer}, {and} \bibinfo{person}{L.~P. Noldus}.} \bibinfo{year}{2009}\natexlab{}.
\newblock \showarticletitle{The {Observer XT}: A tool for the integration and synchronization of multimodal signals}.
\newblock \bibinfo{journal}{\emph{Behav. Res. Methods}} \bibinfo{volume}{41}, \bibinfo{number}{3} (\bibinfo{year}{2009}).
\newblock


\end{thebibliography}
\end{document}